\documentclass[%
superscriptaddress,
notitlepage,
showpacs,
amsmath,amssymb,
aps,
onecolumn,
longbibliography,
]{revtex4-1}

\usepackage{psfrag,graphicx,epsfig,color}
\usepackage{dcolumn}
\usepackage{bm}
\usepackage{natbib}
\usepackage[usenames,dvipsnames,svgnames,table]{xcolor}
\usepackage{subfigure}
\usepackage{rotating}
\usepackage{float}
\usepackage[nice]{nicefrac}

\def\re    {{R_\lambda}}
\def\uu {{\mathbf{u}}}
\def\xx {{\mathbf{x}}}
\def\rr {{\mathbf{r}}}
\def\kk {{\mathbf{k}}}

\begin{document}

\title{
Intermittency of turbulent velocity and scalar fields using 3D local averaging
}

\author{Dhawal Buaria }
\email[]{dhawal.buaria@nyu.edu}
\affiliation{Tandon School of Engineering, New York University, New York, NY 11201, USA}
\affiliation{Max Planck Institute for Dynamics and Self-Organization, 37077 G\"ottingen, Germany}

\author{Katepalli R. Sreenivasan}
\affiliation{Tandon School of Engineering, New York University, New York, NY 11201, USA}
\affiliation{Department of Physics and the Courant Institute of Mathematical Sciences,
New York University, New York, NY 10012, USA}

\date{\today}


\begin{abstract}

An efficient approach for extracting 3D
local averages in spherical subdomains 
is proposed and applied to study the intermittency of small-scale
velocity and scalar fields in direct numerical
simulations of isotropic turbulence. We 
focus on the inertial-range scaling exponents of
locally averaged 
energy dissipation rate, enstrophy 
and scalar dissipation 
rate corresponding to the mixing of a passive scalar $\theta$ in the presence 
of a uniform mean gradient. The Taylor-scale Reynolds number $\re$ goes 
up to $1300$, and the Schmidt number $Sc$ up to $512$ (albeit at smaller $\re$).
The intermittency exponent of the energy dissipation rate
is $\mu \approx 0.23$, whereas that of enstrophy is slightly
larger; trends with $\re$ suggest that this will be the case even at 
extremely large $\re$. The intermittency exponent of the scalar dissipation 
rate is $\mu_\theta \approx 0.35$ for $Sc=1$. 
These findings are in essential agreement with previously reported results in the literature. We further show that $\mu_\theta$ decreases
monotonically with increasing $Sc$, either as $1/\log Sc$ or a weak power law, 
suggesting that
$\mu_\theta \to 0$ as $Sc \to \infty$,  
reaffirming recent results on 
the breakdown of scalar dissipation anomaly
in this limit.

\end{abstract}

\maketitle


\section{Introduction}

A key characteristic of fully developed
fluid turbulence is small-scale
intermittency, referring to the
sporadic generation of intense
fluctuations of velocity gradients
or velocity increments, which result in
strong deviations from Gaussianity and necessitate anomalous corrections
to the seminal mean-field description by Kolmogorov (1941)  \cite{Frisch95, SA97}.
Given its practical importance in numerous
physical processes 
\cite{wilson1996, Falkovich_2002, shaw03,Sreeni04,ham_pof11, BSY.2015}, 
and its fundamental
connection to the energy cascade \cite{Tsi2009},
a quantitative characterization of intermittency
is at the heart of turbulence theory 
\cite{Frisch95, SA97} 
and modeling
\cite{Meneveau11}.
A key concept in understanding intermittency
is the introduction of local averaging 
which allows a quantification of anomalous
corrections to the mean-field description in some pertinent manner \cite{K62,O62}.
In general, for a
fluctuating quantity $A(\xx,t)$,
its local average $A_r(\xx,t)$
over a scale $r$ 
can be defined as:
\begin{align}
A_r (\xx, t) =  \frac{3}{4\pi r^3}  
\int_{|\xx'| \le r} A(\xx + \xx',t) \ d\xx' \ .
\label{eq:lavg}
\end{align}

Evidently, local
averages are defined over a spherical volume
to ensure isotropy with respect to the 
averaging scale size. However, such an averaging
has not been possible until now because the full 3D field is rarely available
in experiments. So they have predominantly
relied on surrogates, which are usually in the form of averages along a line or in a plane
\cite{SA97}. These methods are sometimes known to give different
results compared to 3D local averages \cite{wang1996, sreeni1977, stolo92, iyer2015}.
Even when the full 3D field is available in experiments,
the data are restricted to low Reynolds
numbers \cite{lawson19}, 
where a plausible inertial range is not available.
In contrast, direct numerical simulations (DNS) provide
access to full 3D field at sufficiently large Reynolds numbers, 
but accurate spherical averaging
needs some extra work, since the data are available on a
Cartesian grid. Consequently, recent 
works have relied on 3D averages over cubical
domains \cite{iyer2015, YR2020} 
which, while convenient, could retain some anisotropies.

In this work, we present a simple approach
to efficiently and accurately obtain 3D local averages in 
spherical domains from the DNS data
and apply it to study the intermittency 
of velocity and scalar fields. 
For the velocity field, we revisit inertial-range scaling
of locally averaged energy dissipation rate and enstrophy
\cite{chen97, YR2020}.
For the scalar field, we consider the scaling
of locally averaged scalar dissipation rate,
and compare it to that of the energy dissipation rate.
A key novelty is that we 
focus on mixing of low-diffusivity scalars
(or high Schmidt numbers), which are 
notoriously difficult to obtain
due to additional resolution constraints,
and have been available only very recently
at sufficiently high Reynolds numbers 
\cite{BCSY21a,BCSY21b}.

Our work confirms the past results
obtained primarily in experiments using one or two-dimensional surrogates.
Some important trends with Reynolds numbers are also highlighted.
We also show that the intermittent character of the scalar dissipation disappears at high Schmidt numbers, consistent 
with \cite{BCSY21a,BCSY21b}.

\section{\bf Background}

The energy dissipation rate $\epsilon$ and 
the enstrophy  $\Omega$,
defined, respectively, as
\begin{align}
\epsilon = 2\nu S_{ij} S_{ij}  \ , \ \ \ 
\Omega = \omega_i \omega_i  \ , 
\end{align}
capturing the local
straining and rotational motions,
are central to our understanding of the small-scale
structure of turbulence
\cite{Siggia:1981a, nelkin1999, zeff:2003, BPBY2019, BP2022}.
Here, $\nu$ is the kinematic viscosity,
$S_{ij}$ is the strain-rate tensor and $\omega_i$
is the vorticity
(and repeated indices imply summation).
At high Reynolds numbers, these quantities become highly intermittent, 
so a means to characterizing them,
following Kolmogorov (1962) \cite{K62},
is to average them locally over a scale $r$ and study these averages 
for a wide range of $r$. In particular, it is postulated that for the locally averaged energy dissipation
$\epsilon_r$, its second moment will scale as
\begin{align}
\langle \epsilon_r^2 \rangle \sim r^{-\mu} \ ,
\label{eq:mu}
\end{align}
for $r$ in the inertial range, 
where the constant $\mu$ is termed
as the `intermittency exponent'.
Note, other definitions of intermittency
exponent are also possible, 
but they are mostly equivalent and 
give essentially the same value \cite{sreeni93}; also see Sec. IV~A. 
Based on theoretical grounds \cite{nelkin1999},
a similar result is also anticipated for the locally averaged 
enstrophy $\Omega_r$, with the same numerical value of the intermittency 
exponent. However, previous DNS data have suggested
a slightly larger intermittency exponent for enstrophy
\cite{chen97,YR2020}.

The pertinent
small-scale quantity when considering turbulent
mixing of a passive scalar $\theta (\xx,t)$ 
is the scalar dissipation rate
\begin{align}
\chi = 2 D |\nabla \theta|^2 \ ,
\end{align}
where $D$ is the scalar diffusivity. It is well established that the 
scalar gradients and scalar increments
also exhibit intermittency \cite{SA97}, and so, similar to $\epsilon_r$ and
$\Omega_r$, we can consider the
scaling of $\chi_r$ \cite{prasad88}. The mixing process is controlled additionally by the Schmidt number $Sc = \nu/D$.
Obtaining data at high $Sc$, while also keeping the Reynolds numbers acceptably high,
is extremely challenging due to additional resolution constraints, 
and has only been possible very recently 
\cite{BCSY21a,BCSY21b}.

\section{Numerical Approach}

\subsection{Direct numerical simulations and database}

The DNS data examined in this work
are obtained by solving the incompressible
Navier-Stokes equations:
\begin{align}
\partial \uu /\partial t +  \uu\cdot\nabla \uu
= - \nabla P + \nu \nabla^2 \uu + \mathbf{f} \ ,
\end{align}
where $\uu$ is the divergence free
velocity field ($\nabla \cdot \uu=0$),
$P$ is the kinematic pressure and $\mathbf{f}$
is the large-scale forcing term to maintain statistical
stationarity. The DNS corresponds to the canonical 
setup of isotropic turbulence 
in a periodic domain \cite{Ishihara09}, 
allowing the use of highly accurate Fourier pseudo-spectral 
methods, with aliasing errors controlled
using a combination of grid-shifting
and truncation \cite{Rogallo}.
The database for the present work corresponds to recent works
\cite{BS2020, BBP2020, BPB2022, BS2022},
with the Taylor-scale Reynolds number 
$\re$ in the range $140-1300$. Convergence with respect to 
resolution and statistical
sampling has also been thoroughly established in all 
these previous works.

The passive scalar is obtained by simultaneously 
solving the advection-diffusion equation in the presence of mean uniform gradient:
\begin{align}
\partial \theta /\partial t  + \uu \cdot \nabla \theta
 = - \uu \cdot \nabla \Theta + D \nabla^2 \theta \ .
\end{align}
The uniform mean gradient is set as
$\nabla \Theta = (G,0,0)$ along the first Cartesian directions,
and provides the forcing needed to achieve statistical
stationarity for the scalar \cite{overholt96}.
The database for scalars utilized here is the same as in our recent papers \cite{BCSY21a,BCSY21b}, and corresponds to  $\re$
in the range $140-650$, and $Sc$ in the range $1-512$.
As noted in \cite{BCSY21a}, the data were generated using 
conventional Fourier pseudo-spectral 
methods for $Sc=1$, and a hybrid approach for 
higher $Sc$ \cite{clay.cpc1, clay.cpc2,clay.omp}; this approach consisted of 
solving the velocity field pseudo-spectrally 
while resolving the Kolmogorov length scale $\eta_K$, while 
resolving the scalar field using compact finite differences 
on a finer grid, so as to resolve 
the Batchelor scale $\eta_B =  \eta_K Sc^{-1/2}$ \cite{batch1959a}.

\subsection{Local averaging procedure}

To implement the 3D local averaging efficiently, Eq.~\eqref{eq:lavg} is rewritten as:
\begin{align}
A_r (\xx, t) =  \int_{\xx'} G(\xx') A(\xx + \xx',t) \ d\xx',
\end{align}
where $G(\xx') = 3/4\pi r^3$, 
satisfying $\int_{\xx'} G(\xx') d\xx' = 1$,
represents an isotropic box filter \cite{popebook}. 
Following recent work in Ref.~\cite{BPB2020, BP2021}, this filtering operation can 
be easily evaluated in the Fourier-space 
for a chosen value of $r$, as:
\begin{align}
\hat{A}_r (\kk, t)  = f(kr) \hat{A} (\kk, t) \ , 
\ \ \ \text{where} \ \ \ \ 
f(kr) = \frac{3 \left[ \sin(kr) - kr \cos(kr) \right] }{(kr)^3} \  .
\end{align}
Here, $\hat{(\cdot)}$ denotes the Fourier transform,
$\mathbf{k}$ is the wave-vector, with $|\mathbf{k}|=k$,
and $f(kr)$ is the transfer function corresponding to $G(\rr)$. 
Unlike in previous works \cite{iyer2015,YR2020}, this local averaging is evaluated
exactly in an isotropic spherical volume.
It can also be easily shown that this isotropic box filter 
also satisfies the consistency condition: $\langle A_r \rangle = \langle A \rangle$.

\section{Results}

\begin{figure}
\begin{center}
\includegraphics[width=16.0cm]{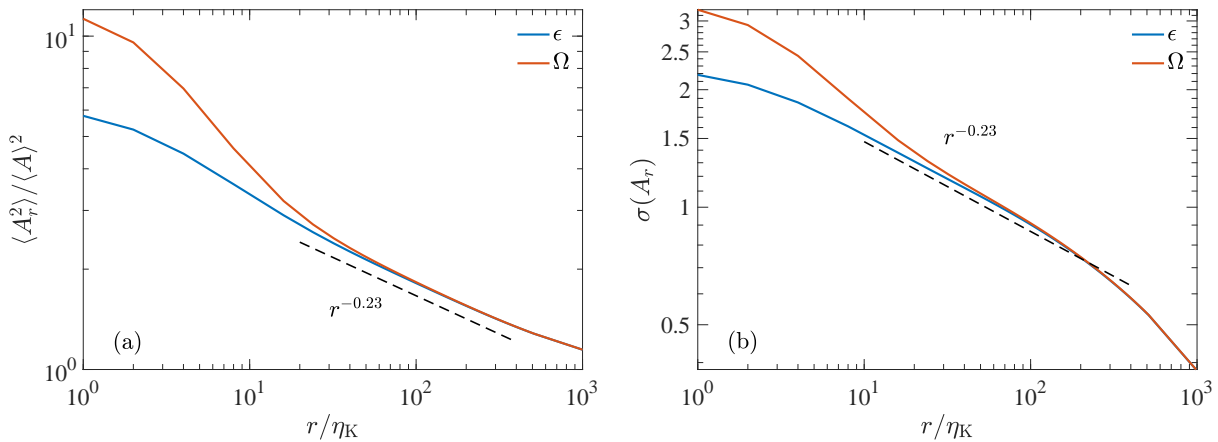}
\caption{
(a) Second moment of locally averaged dissipation and
enstrophy at $\re=1300$. The intermittency exponent
for dissipation is about $\mu=0.23$
and that for enstrophy is slightly larger. 
(b) The variance of local averages as defined
by Eq.~\eqref{eq:varl}.
}
\label{fig:l1300}
\end{center}
\end{figure}

\subsection{Energy dissipation rate and enstrophy}

We first consider the intermittency
of dissipation and enstrophy,
a topic that has received considerable
attention in the literature 
\cite{Siggia:1981a, chen97, grossmann1997, BPBY2019, YR2020}.
Figure~\ref{fig:l1300}a shows
the second moments of locally-averaged
dissipation and enstrophy 
at $\re=1300$ (the highest in the current work).
As anticipated, for small $r$, the moments of local averages 
simply tend to those of instantaneous quantities,
those for enstrophy being larger (as is known); whereas at large $r$, 
the moments tend to the same value of unity (as they should).
In the inertial range, 
dissipation exhibits the power-law 
$r^{-0.23}$ (with an error bar of $0.02$) for a reasonable range of $r$,
implying $\mu\approx0.23\pm0.02$, in very good
agreement with previous works \cite{sreeni93}.
At the same time, within the same range of $r$,
the enstrophy curve suggests a slightly
larger intermittency exponent, consistent with 
previous works \cite{chen97,YR2020}; see later for more details. 

An alternative means of extracting the intermittency exponent is to 
consider the variance of $\epsilon_r$ (and $\Omega_r$), computed after subtracting the respective means for each $r$; for instance,
\begin{align}
\sigma(\epsilon_r) = 
\langle (\epsilon_r - \langle \epsilon_r \rangle)^2 \rangle / 
\langle \epsilon_r \rangle^2 = 
\langle \epsilon_r^2 \rangle / \langle \epsilon \rangle^2 - 1 \ ;
\label{eq:varl}
\end{align}
note that $\langle \epsilon_r \rangle = \langle \epsilon \rangle$.
The expectation is that $\sigma(\epsilon_r) \sim r^{-\mu}$ 
in the inertial range \cite{sreeni93}. 
Figure~\ref{fig:l1300}b shows the variance of
$\epsilon_r$ and $\Omega_r$, and indeed the nature of their scaling
follow expectations --- 
although the scaling ranges are somewhat different from Fig.~\ref{fig:l1300}a.

\begin{figure}
\begin{center}
\includegraphics[width=16.0cm]{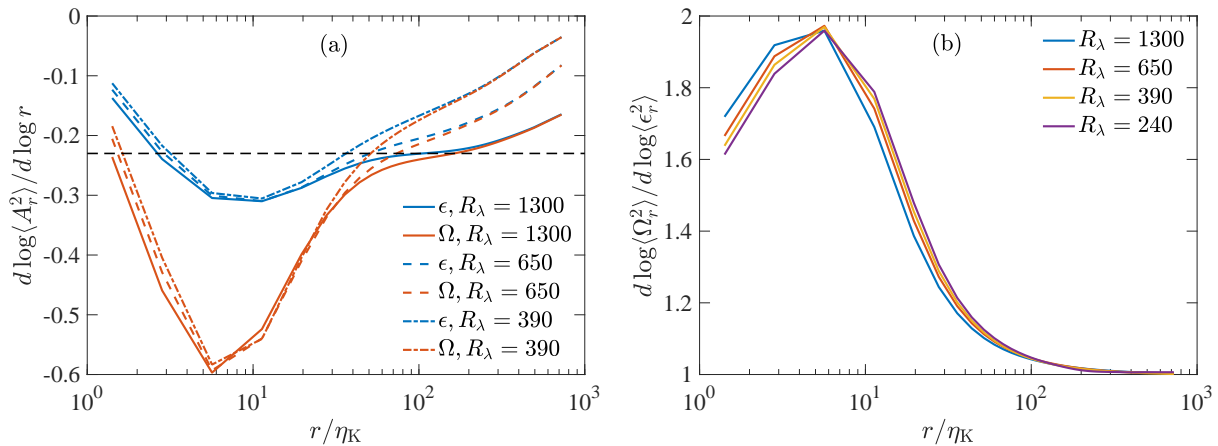}
\caption{
(a) Local slopes of the second moment of locally averaged
dissipation and enstrophy at different Reynolds numbers. The constancy of slopes improves as the inertial range enlarges with increasing Reynolds numbers. (b) Ratio of the two local slopes as a function of $r$. It is evident that the intermittency exponent
for enstrophy is slightly larger and the approach towards
unity as Reynolds number increases is extremely slow.
}
\label{fig:lre}
\end{center}
\end{figure}

\subsection{Effect of Reynolds number}
To be stringent about the power law exponents, it is helpful to take the
log-log derivatives (or the local slope) of the curves in 
Fig.~\ref{fig:l1300}a. Figure~\ref{fig:lre}a shows
the local slope of second moments 
of $\epsilon_r$ and $\Omega_r$ for various $\re$.
It can be seen that the quality of results
in the inertial range depend
on $\re$ and, in fact, a constant local
slope (corresponding to a true power-law)
is convincing only at the 
highest $\re$ ($=1300$). However, guided by this feature, one can look for signs of approximate power laws at lower $\re$, and find slightly large exponent values. 

It is worth noting that
in Fig.~\ref{fig:lre}a, 
the local slope of enstrophy
(in the inertial range) is always larger
than that of dissipation for every $\re$.
To better document this behavior,
Fig.~\ref{fig:lre}b shows the ratio of the local slope
of enstrophy with respect to that
of dissipation (in the spirit of 
extended self-similarity). Remarkably, the curves
are always above unity in the inertial range.
Further, there is a weak but clear tendency for this ratio to approach unity with 
increasing $\re$, but the rate of approach is so slow that the difference remains 
in place at all finite Reynolds numbers of interest. That is, 
enstrophy in the inertial range
is slightly more intermittent than dissipation. 

\subsection{Scalar dissipation rate}

\begin{figure}
\begin{center}
\includegraphics[width=16.0cm]{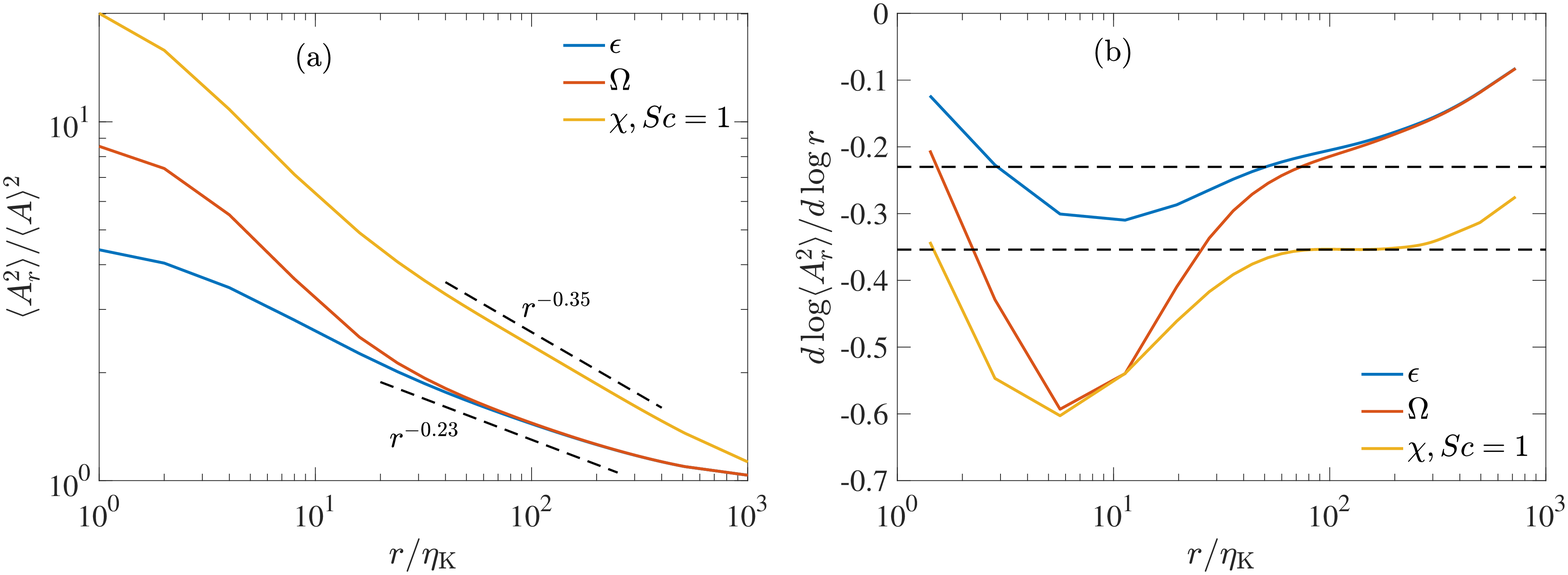}
\caption{
(a) Second moment of locally averaged 
energy dissipation, enstrophy and scalar dissipation 
at $\re=650$, with scalar at $Sc=1$.
The intermittency exponent of scalar
$\mu=0.35$ is significantly larger.
(b) The local slopes of quantities shown in (a) demonstrates
that the inertial range scaling for the scalar is very robust.
}
\label{fig:l650}
\end{center}
\end{figure}

We first consider locally averaged scalar dissipation
rate for $Sc=1$. 
Figure~\ref{fig:l650}a shows the 
second moments of locally averaged
energy dissipation, enstrophy and scalar dissipation
(at $\re=650$), with the corresponding local slopes shown in
Fig.~\ref{fig:l650}b.
The intermittency exponent
of the scalar $\mu_\theta$ is
larger than those of both dissipation
and enstrophy. While 
a clear plateau for dissipation
and enstrophy is not achieved
at $\re=650$ (in Fig.~\ref{fig:l650}b)
the curve for scalar dissipation
shows a distinct plateau,
giving $\mu_\theta \approx 0.35$.
This result is in excellent agreement
with earlier experimental results 
of \cite{sreeni1977,prasad88}.

\begin{figure}
\begin{center}
\includegraphics[width=16.0cm]{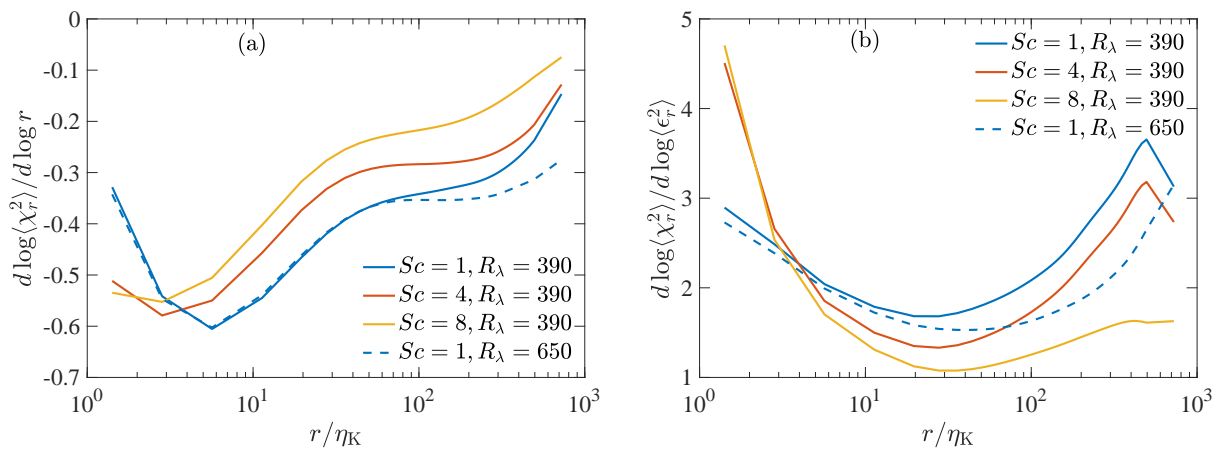}
\caption{
(a) Local slopes of the second moment of locally averaged 
scalar dissipation at $\re=390$ for $Sc=1,4,8$ and at
$\re=650$ for $Sc=1$. 
(b) The ratio of the local slope of scalar dissipation to that of
energy dissipation for same cases as shown in (a). 
}
\label{fig:l390}
\end{center}
\end{figure}

The effect of increasing $Sc$ is considered
next. Since increasing $Sc$ imposes
a stricter constraint on small-scale
resolution, we consider data
at slightly lower $\re=390$,
but still large enough to display
inertial range characteristics.
Figure~\ref{fig:l390}a shows the local
slope of second moment of scalar dissipation
at $\re=390$ and $Sc=1-8$. 
The curve corresponding to $\re=650$ and $Sc=1$ 
is also shown for comparison.
The scalar intermittency exponent monotonically
decreases with increasing $Sc$ 
(also displaying a weak $\re$ dependence). 

Due to this possible $\re$-dependence,
it is difficult to
extract precisely the intermittency exponents
at higher $Sc$. Instead, 
if we were to compare the ratio of
$\mu_\theta$ to $\mu$,
a trend can be established; this
should shed light
on the asymptotic limit of $Sc\to\infty$. 
To this end,
Fig.~\ref{fig:l390}b shows
the local slope of scalar dissipation
with respect to that of energy dissipation.
The $\re$-dependence is somewhat more
prominent than in Fig.~\ref{fig:l390}a,
likely because $\mu_\theta$ seemingly has a stronger
$\re$-dependence than $\mu$. 
Nevertheless, it is evident
that the ratio $\mu_\theta/\mu$ monotonically decreases with $Sc$.

\begin{figure}
\begin{center}
\includegraphics[width=16.0cm]{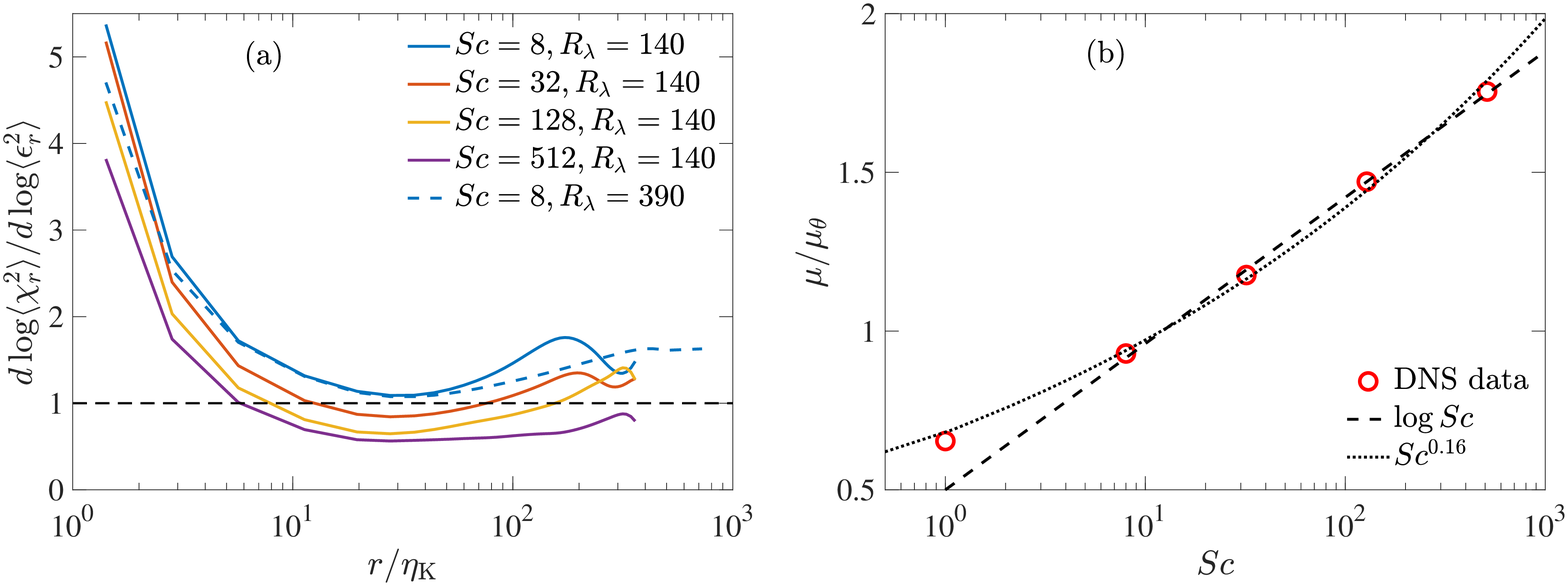}
\caption{
(a) Ratio of local slopes of the second moments of locally averaged 
scalar dissipation and energy dissipation at $\re=140$ for $Sc=8,32,128,512$ and at
$\re=390$ for $Sc=8$. 
(b) Inverse of the inertial range exponents extracted from (a)
as a function of $Sc$. Both power-law and log fits are shown.
}
\label{fig:l140}
\end{center}
\end{figure}

For a definitive answer on the high $Sc$ limit, one obviously needs
to obtain data at substantially higher $Sc$
at $\re=390$ (and also at higher $\re$).
But these are unlikely to be attained anytime soon.
Instead, we analyze data at lower $\re=140$,
for which inertial range characteristics just
begin to manifest \cite{Ishihara09,iyer2015}.
Figure~\ref{fig:l140}a shows the
local slope of second moment of scalar
dissipation with respect to that of
energy dissipation up to $Sc$ as high as $512$.
The curve corresponding to $\re=390$ and $Sc=8$
is also shown for comparison.

Two main conclusions can be drawn
from this figure. First,
the effect of $\re$ is 
weaker at higher $Sc$
(in comparison to that at $Sc=1$,
as evident from Fig.~\ref{fig:l390}b).
Second, the ratio $\mu_\theta/\mu$
keeps decreasing further, 
possibly suggesting that 
$\mu_\theta \to 0$ as $Sc \to \infty$.
We extract the ratio $\mu_\theta/\mu$
from Fig.~\ref{fig:l140}a and
plot its inverse as a function of $Sc$ in 
Fig.~\ref{fig:l140}b. 
The data show a weak power-law
dependence (with an exponent of about $0.16$), though a $\log Sc$-behavior 
is equally appropriate. The 
$\log Sc$-behavior of $\mu/\mu_\theta$ is loosely
based on how the mean scalar dissipation rate 
scales with $Sc$ \cite{BCSY21b}.

\section{Conclusions}

In this work, we have performed local averaging of turbulent intermittent variables 
as they should be done: spherical averaging of three-dimensional 
quantities without using surrogates from one- or two-dimensional cuts, 
without the use of Taylor's frozen-flow hypothesis, or relying on 
cubical subdomains. For the energy 
dissipation and enstrophy, we mostly recover past results with greater assurance; we feel more confident than 
ever that intermittency exponents do exist. Our results additionally suggest that enstrophy will always 
remain somewhat more intermittent than energy 
dissipation even at extremely large $\re$, 
reaffirming recent results of \cite{BP2022}. 

For the scalar dissipation, 
no previous studies existed on its scaling at high Schmidt numbers with large enough Reynolds number to obtain inertial range characteristics. 
In our simulations, the maximum allowable computational capacity has necessitated 
decreasing $Sc$ with increasing $\re$. 
For $Sc = 1$, we were able to maintain $\re$ as high as 650, whereas $\re =140$
for the highest $Sc=512$. It is unlikely that one can get much higher $\re$ for such high $Sc$ anytime soon. For these conditions, we have been able to extract the intermittency exponent for the scalar dissipation. For $Sc = 1$, 
it is very close to what had been deduced from earlier measurements of the full 
scalar dissipation using Taylor's hypothesis \cite{sreeni1977}, and some surrogate quantities in planar cuts \cite{prasad88}. An important 
result is that the intermittency exponent for the scalar dissipation decreases 
either as a weak power of $Sc$ or logarithmically, apparently to zero in the limit
$Sc \to \infty$. This decrease is consistent with our recent results 
\cite{BCSY21a,BCSY21b} that turbulence loses its ability to effectively 
mix passive scalars with very low diffusivity.

\vspace{0.25cm}
\paragraph*{Acknowledgments:} 

We thank P.K. Yeung for their comments on the draft 
and sustained collaboration over the years.
D.B. gratefully acknowledge the Gauss Centre
for Supercomputing e.V. (www.gauss-centre.eu) for providing computing time
on the supercomputers JUQUEEN and JUWELS at J\"ulich Supercomputing Centre,
where the simulations utilized in this work were primarily performed.
The high Schmidt number simulations at $\re=140$ ($Sc\ge8$) and 
$\re=390$ ($Sc=8$) were 
performed together with Matthew Clay and P.K. Yeung 
using resources of the Oak Ridge 
Leadership Computing Facility (OLCF), under
2017 and 2018 INCITE Awards.



\begin{thebibliography}{43}%
\makeatletter
\providecommand \@ifxundefined [1]{%
 \@ifx{#1\undefined}
}%
\providecommand \@ifnum [1]{%
 \ifnum #1\expandafter \@firstoftwo
 \else \expandafter \@secondoftwo
 \fi
}%
\providecommand \@ifx [1]{%
 \ifx #1\expandafter \@firstoftwo
 \else \expandafter \@secondoftwo
 \fi
}%
\providecommand \natexlab [1]{#1}%
\providecommand \enquote  [1]{``#1''}%
\providecommand \bibnamefont  [1]{#1}%
\providecommand \bibfnamefont [1]{#1}%
\providecommand \citenamefont [1]{#1}%
\providecommand \href@noop [0]{\@secondoftwo}%
\providecommand \href [0]{\begingroup \@sanitize@url \@href}%
\providecommand \@href[1]{\@@startlink{#1}\@@href}%
\providecommand \@@href[1]{\endgroup#1\@@endlink}%
\providecommand \@sanitize@url [0]{\catcode `\\12\catcode `\$12\catcode
  `\&12\catcode `\#12\catcode `\^12\catcode `\_12\catcode `\%12\relax}%
\providecommand \@@startlink[1]{}%
\providecommand \@@endlink[0]{}%
\providecommand \url  [0]{\begingroup\@sanitize@url \@url }%
\providecommand \@url [1]{\endgroup\@href {#1}{\urlprefix }}%
\providecommand \urlprefix  [0]{URL }%
\providecommand \Eprint [0]{\href }%
\providecommand \doibase [0]{http://dx.doi.org/}%
\providecommand \selectlanguage [0]{\@gobble}%
\providecommand \bibinfo  [0]{\@secondoftwo}%
\providecommand \bibfield  [0]{\@secondoftwo}%
\providecommand \translation [1]{[#1]}%
\providecommand \BibitemOpen [0]{}%
\providecommand \bibitemStop [0]{}%
\providecommand \bibitemNoStop [0]{.\EOS\space}%
\providecommand \EOS [0]{\spacefactor3000\relax}%
\providecommand \BibitemShut  [1]{\csname bibitem#1\endcsname}%
\let\auto@bib@innerbib\@empty
\bibitem [{\citenamefont {Frisch}(1995)}]{Frisch95}%
  \BibitemOpen
  \bibfield  {author} {\bibinfo {author} {\bibfnamefont {U.}~\bibnamefont
  {Frisch}},\ }\href@noop {} {\emph {\bibinfo {title} {Turbulence: the legacy
  of {Kolmogorov}}}}\ (\bibinfo  {publisher} {Cambridge University Press},\
  \bibinfo {address} {Cambridge},\ \bibinfo {year} {1995})\BibitemShut
  {NoStop}%
\bibitem [{\citenamefont {Sreenivasan}\ and\ \citenamefont
  {Antonia}(1997)}]{SA97}%
  \BibitemOpen
  \bibfield  {author} {\bibinfo {author} {\bibfnamefont {K.~S.}\ \bibnamefont
  {Sreenivasan}}\ and\ \bibinfo {author} {\bibfnamefont {R.~A.}\ \bibnamefont
  {Antonia}},\ }\bibfield  {title} {\enquote {\bibinfo {title} {The
  phenomenology of small-scale turbulence},}\ }\href@noop {} {\bibfield
  {journal} {\bibinfo  {journal} {Annu.~Rev.~Fluid~Mech.}\ }\textbf {\bibinfo
  {volume} {29}},\ \bibinfo {pages} {435--77} (\bibinfo {year}
  {1997})}\BibitemShut {NoStop}%
\bibitem [{\citenamefont {Wilson}\ \emph {et~al.}(1996)\citenamefont {Wilson},
  \citenamefont {Wyngaard},\ and\ \citenamefont {Havelock}}]{wilson1996}%
  \BibitemOpen
  \bibfield  {author} {\bibinfo {author} {\bibfnamefont {D.~K.}\ \bibnamefont
  {Wilson}}, \bibinfo {author} {\bibfnamefont {J.~C.}\ \bibnamefont
  {Wyngaard}}, \ and\ \bibinfo {author} {\bibfnamefont {D.~I.}\ \bibnamefont
  {Havelock}},\ }\bibfield  {title} {\enquote {\bibinfo {title} {The effect of
  turbulent intermittency on scattering into an acoustic shadow zone},}\
  }\href@noop {} {\bibfield  {journal} {\bibinfo  {journal} {J. Acoust. Soc.
  Am.}\ }\textbf {\bibinfo {volume} {99}},\ \bibinfo {pages} {3393--3400}
  (\bibinfo {year} {1996})}\BibitemShut {NoStop}%
\bibitem [{\citenamefont {Falkovich}\ \emph {et~al.}(2002)\citenamefont
  {Falkovich}, \citenamefont {Fouxon},\ and\ \citenamefont
  {Stepanov}}]{Falkovich_2002}%
  \BibitemOpen
  \bibfield  {author} {\bibinfo {author} {\bibfnamefont {G.}~\bibnamefont
  {Falkovich}}, \bibinfo {author} {\bibfnamefont {A.}~\bibnamefont {Fouxon}}, \
  and\ \bibinfo {author} {\bibfnamefont {M.~G.}\ \bibnamefont {Stepanov}},\
  }\bibfield  {title} {\enquote {\bibinfo {title} {Acceleration of rain
  initiation by cloud turbulence},}\ }\href@noop {} {\bibfield  {journal}
  {\bibinfo  {journal} {Nature}\ }\textbf {\bibinfo {volume} {419}},\ \bibinfo
  {pages} {151} (\bibinfo {year} {2002})}\BibitemShut {NoStop}%
\bibitem [{\citenamefont {Shaw}(2003)}]{shaw03}%
  \BibitemOpen
  \bibfield  {author} {\bibinfo {author} {\bibfnamefont {R.~A.}\ \bibnamefont
  {Shaw}},\ }\bibfield  {title} {\enquote {\bibinfo {title}
  {Particle-turbulence interactions in atmospheric clouds},}\ }\href@noop {}
  {\bibfield  {journal} {\bibinfo  {journal} {Annu.~Rev.~Fluid Mech.}\ }\textbf
  {\bibinfo {volume} {35}},\ \bibinfo {pages} {183--227} (\bibinfo {year}
  {2003})}\BibitemShut {NoStop}%
\bibitem [{\citenamefont {Sreenivasan}(2004)}]{Sreeni04}%
  \BibitemOpen
  \bibfield  {author} {\bibinfo {author} {\bibfnamefont {K.~R.}\ \bibnamefont
  {Sreenivasan}},\ }\bibfield  {title} {\enquote {\bibinfo {title} {Possible
  effects of small-scale intermittency in turbulent reacting flows},}\
  }\href@noop {} {\bibfield  {journal} {\bibinfo  {journal} {Flow, Turb.
  Comb.}\ }\textbf {\bibinfo {volume} {72}},\ \bibinfo {pages} {115--131}
  (\bibinfo {year} {2004})}\BibitemShut {NoStop}%
\bibitem [{\citenamefont {Hamlington}\ \emph {et~al.}(2011)\citenamefont
  {Hamlington}, \citenamefont {Poludnenko},\ and\ \citenamefont
  {Oran}}]{ham_pof11}%
  \BibitemOpen
  \bibfield  {author} {\bibinfo {author} {\bibfnamefont {P.~E.}\ \bibnamefont
  {Hamlington}}, \bibinfo {author} {\bibfnamefont {A.~Y.}\ \bibnamefont
  {Poludnenko}}, \ and\ \bibinfo {author} {\bibfnamefont {E.~S.}\ \bibnamefont
  {Oran}},\ }\bibfield  {title} {\enquote {\bibinfo {title} {Interactions
  between turbulence and flames in premixed reacting flows},}\ }\href@noop {}
  {\bibfield  {journal} {\bibinfo  {journal} {Physics of Fluids}\ }\textbf
  {\bibinfo {volume} {23}},\ \bibinfo {pages} {125111} (\bibinfo {year}
  {2011})}\BibitemShut {NoStop}%
\bibitem [{\citenamefont {Buaria}\ \emph {et~al.}(2015)\citenamefont {Buaria},
  \citenamefont {Sawford},\ and\ \citenamefont {Yeung}}]{BSY.2015}%
  \BibitemOpen
  \bibfield  {author} {\bibinfo {author} {\bibfnamefont {D.}~\bibnamefont
  {Buaria}}, \bibinfo {author} {\bibfnamefont {B.~L.}\ \bibnamefont {Sawford}},
  \ and\ \bibinfo {author} {\bibfnamefont {P.~K.}\ \bibnamefont {Yeung}},\
  }\bibfield  {title} {\enquote {\bibinfo {title} {Characteristics of backward
  and forward two-particle relative dispersion in turbulence at different
  {R}eynolds numbers},}\ }\href@noop {} {\bibfield  {journal} {\bibinfo
  {journal} {Phys. Fluids}\ }\textbf {\bibinfo {volume} {27}},\ \bibinfo
  {pages} {105101} (\bibinfo {year} {2015})}\BibitemShut {NoStop}%
\bibitem [{\citenamefont {Tsinober}(2009)}]{Tsi2009}%
  \BibitemOpen
  \bibfield  {author} {\bibinfo {author} {\bibfnamefont {A.}~\bibnamefont
  {Tsinober}},\ }\href@noop {} {\emph {\bibinfo {title} {An Informal Conceptual
  Introduction to Turbulence}}}\ (\bibinfo  {publisher} {Springer},\ \bibinfo
  {address} {Berlin},\ \bibinfo {year} {2009})\BibitemShut {NoStop}%
\bibitem [{\citenamefont {Meneveau}(2011)}]{Meneveau11}%
  \BibitemOpen
  \bibfield  {author} {\bibinfo {author} {\bibfnamefont {C.}~\bibnamefont
  {Meneveau}},\ }\bibfield  {title} {\enquote {\bibinfo {title} {Lagrangian
  dynamics and models of the velocity gradient tensor in turbulent flows},}\
  }\href@noop {} {\bibfield  {journal} {\bibinfo  {journal} {Annu. Rev. Fluid
  Mech.}\ }\textbf {\bibinfo {volume} {43}},\ \bibinfo {pages} {219--245}
  (\bibinfo {year} {2011})}\BibitemShut {NoStop}%
\bibitem [{\citenamefont {Kolmogorov}(1962)}]{K62}%
  \BibitemOpen
  \bibfield  {author} {\bibinfo {author} {\bibfnamefont {A.~N.}\ \bibnamefont
  {Kolmogorov}},\ }\bibfield  {title} {\enquote {\bibinfo {title} {A refinement
  of previous hypotheses concerning the local structure of turbulence in a
  viscous incompressible fluid at high {Reynolds} number},}\ }\href@noop {}
  {\bibfield  {journal} {\bibinfo  {journal} {J.~Fluid Mech.}\ }\textbf
  {\bibinfo {volume} {13}},\ \bibinfo {pages} {82--85} (\bibinfo {year}
  {1962})}\BibitemShut {NoStop}%
\bibitem [{\citenamefont {Oboukhov}(1962)}]{O62}%
  \BibitemOpen
  \bibfield  {author} {\bibinfo {author} {\bibfnamefont {A.~M.}\ \bibnamefont
  {Oboukhov}},\ }\bibfield  {title} {\enquote {\bibinfo {title} {Some specific
  features of atmospheric tubulence},}\ }\href@noop {} {\bibfield  {journal}
  {\bibinfo  {journal} {J.~Fluid Mech.}\ }\textbf {\bibinfo {volume} {13}},\
  \bibinfo {pages} {77--81} (\bibinfo {year} {1962})}\BibitemShut {NoStop}%
\bibitem [{\citenamefont {Wang}\ \emph {et~al.}(1996)\citenamefont {Wang},
  \citenamefont {Chen}, \citenamefont {Brasseur},\ and\ \citenamefont
  {Wyngaard}}]{wang1996}%
  \BibitemOpen
  \bibfield  {author} {\bibinfo {author} {\bibfnamefont {L.-P.}\ \bibnamefont
  {Wang}}, \bibinfo {author} {\bibfnamefont {S.}~\bibnamefont {Chen}}, \bibinfo
  {author} {\bibfnamefont {J.~G.}\ \bibnamefont {Brasseur}}, \ and\ \bibinfo
  {author} {\bibfnamefont {J.~C.}\ \bibnamefont {Wyngaard}},\ }\bibfield
  {title} {\enquote {\bibinfo {title} {Examination of hypotheses in the
  {Kolmogorov} refined turbulence theory through high-resolution simulations.
  part 1. velocity field},}\ }\href@noop {} {\bibfield  {journal} {\bibinfo
  {journal} {J.~Fluid Mech.}\ }\textbf {\bibinfo {volume} {309}},\ \bibinfo
  {pages} {113--156} (\bibinfo {year} {1996})}\BibitemShut {NoStop}%
\bibitem [{\citenamefont {Sreenivasan}\ \emph {et~al.}(1997)\citenamefont
  {Sreenivasan}, \citenamefont {Antonia},\ and\ \citenamefont
  {Danh}}]{sreeni1977}%
  \BibitemOpen
  \bibfield  {author} {\bibinfo {author} {\bibfnamefont {K.R.}\ \bibnamefont
  {Sreenivasan}}, \bibinfo {author} {\bibfnamefont {R.A.}\ \bibnamefont
  {Antonia}}, \ and\ \bibinfo {author} {\bibfnamefont {H.Q.}\ \bibnamefont
  {Danh}},\ }\bibfield  {title} {\enquote {\bibinfo {title} {Log-normality of
  temperature dissipation in a turbulent boundary layer},}\ }\href@noop {}
  {\bibfield  {journal} {\bibinfo  {journal} {Phys. Fluids}\ }\textbf {\bibinfo
  {volume} {20}},\ \bibinfo {pages} {1238--1249} (\bibinfo {year}
  {1997})}\BibitemShut {NoStop}%
\bibitem [{\citenamefont {Stolovitzky}\ \emph {et~al.}(1992)\citenamefont
  {Stolovitzky}, \citenamefont {Kailasnath},\ and\ \citenamefont
  {Sreenivasan}}]{stolo92}%
  \BibitemOpen
  \bibfield  {author} {\bibinfo {author} {\bibfnamefont {G.}~\bibnamefont
  {Stolovitzky}}, \bibinfo {author} {\bibfnamefont {P.}~\bibnamefont
  {Kailasnath}}, \ and\ \bibinfo {author} {\bibfnamefont {K.~R.}\ \bibnamefont
  {Sreenivasan}},\ }\bibfield  {title} {\enquote {\bibinfo {title}
  {{Kolmogorov’s} refined similarity hypotheses},}\ }\href@noop {} {\bibfield
   {journal} {\bibinfo  {journal} {Phys.~Rev.~Lett.}\ }\textbf {\bibinfo
  {volume} {69}},\ \bibinfo {pages} {1178} (\bibinfo {year}
  {1992})}\BibitemShut {NoStop}%
\bibitem [{\citenamefont {Iyer}\ \emph {et~al.}(2015)\citenamefont {Iyer},
  \citenamefont {Sreenivasan},\ and\ \citenamefont {Yeung}}]{iyer2015}%
  \BibitemOpen
  \bibfield  {author} {\bibinfo {author} {\bibfnamefont {K.~P.}\ \bibnamefont
  {Iyer}}, \bibinfo {author} {\bibfnamefont {K.~R.}\ \bibnamefont
  {Sreenivasan}}, \ and\ \bibinfo {author} {\bibfnamefont {P.~K.}\ \bibnamefont
  {Yeung}},\ }\bibfield  {title} {\enquote {\bibinfo {title} {Refined
  similarity hypothesis using three-dimensional local averages},}\ }\href@noop
  {} {\bibfield  {journal} {\bibinfo  {journal} {Phys.~Rev.~E}\ }\textbf
  {\bibinfo {volume} {92}},\ \bibinfo {pages} {063024} (\bibinfo {year}
  {2015})}\BibitemShut {NoStop}%
\bibitem [{\citenamefont {Lawson}\ \emph {et~al.}(2019)\citenamefont {Lawson},
  \citenamefont {Bodenschatz}, \citenamefont {Knutsen}, \citenamefont
  {Dawson},\ and\ \citenamefont {Worth}}]{lawson19}%
  \BibitemOpen
  \bibfield  {author} {\bibinfo {author} {\bibfnamefont {J.~M.}\ \bibnamefont
  {Lawson}}, \bibinfo {author} {\bibfnamefont {E.}~\bibnamefont {Bodenschatz}},
  \bibinfo {author} {\bibfnamefont {A.~N.}\ \bibnamefont {Knutsen}}, \bibinfo
  {author} {\bibfnamefont {J.~R.}\ \bibnamefont {Dawson}}, \ and\ \bibinfo
  {author} {\bibfnamefont {N.~A.}\ \bibnamefont {Worth}},\ }\bibfield  {title}
  {\enquote {\bibinfo {title} {Direct assessment of {Kolmogorov's} first
  refined similarity hypothesis},}\ }\href@noop {} {\bibfield  {journal}
  {\bibinfo  {journal} {Phys.~Rev.~Fluids}\ }\textbf {\bibinfo {volume} {4}},\
  \bibinfo {pages} {022601} (\bibinfo {year} {2019})}\BibitemShut {NoStop}%
\bibitem [{\citenamefont {Yeung}\ and\ \citenamefont
  {Ravikumar}(2020)}]{YR2020}%
  \BibitemOpen
  \bibfield  {author} {\bibinfo {author} {\bibfnamefont {P.~K.}\ \bibnamefont
  {Yeung}}\ and\ \bibinfo {author} {\bibfnamefont {K.}~\bibnamefont
  {Ravikumar}},\ }\bibfield  {title} {\enquote {\bibinfo {title} {Advancing
  understanding of turbulence through extreme-scale computation: Intermittency
  and simulations at large problem sizes},}\ }\href@noop {} {\bibfield
  {journal} {\bibinfo  {journal} {Phys.~Rev.~Fluids}\ }\textbf {\bibinfo
  {volume} {5}},\ \bibinfo {pages} {110517} (\bibinfo {year}
  {2020})}\BibitemShut {NoStop}%
\bibitem [{\citenamefont {Chen}\ \emph {et~al.}(1997)\citenamefont {Chen},
  \citenamefont {Sreenivasan},\ and\ \citenamefont {Nelkin}}]{chen97}%
  \BibitemOpen
  \bibfield  {author} {\bibinfo {author} {\bibfnamefont {S.}~\bibnamefont
  {Chen}}, \bibinfo {author} {\bibfnamefont {K.~R.}\ \bibnamefont
  {Sreenivasan}}, \ and\ \bibinfo {author} {\bibfnamefont {M.}~\bibnamefont
  {Nelkin}},\ }\bibfield  {title} {\enquote {\bibinfo {title} {Inertial range
  scalings of dissipation and enstrophy in isotropic turbulence},}\ }\href@noop
  {} {\bibfield  {journal} {\bibinfo  {journal} {Phys.~Rev.~Lett.}\ }\textbf
  {\bibinfo {volume} {79}},\ \bibinfo {pages} {1253} (\bibinfo {year}
  {1997})}\BibitemShut {NoStop}%
\bibitem [{\citenamefont {Buaria}\ \emph
  {et~al.}(2021{\natexlab{a}})\citenamefont {Buaria}, \citenamefont {Clay},
  \citenamefont {Sreenivasan},\ and\ \citenamefont {Yeung}}]{BCSY21a}%
  \BibitemOpen
  \bibfield  {author} {\bibinfo {author} {\bibfnamefont {D.}~\bibnamefont
  {Buaria}}, \bibinfo {author} {\bibfnamefont {M.~P.}\ \bibnamefont {Clay}},
  \bibinfo {author} {\bibfnamefont {K.~R.}\ \bibnamefont {Sreenivasan}}, \ and\
  \bibinfo {author} {\bibfnamefont {P.~K.}\ \bibnamefont {Yeung}},\ }\bibfield
  {title} {\enquote {\bibinfo {title} {Turbulence is an ineffective mixer when
  schmidt numbers are large},}\ }\href@noop {} {\bibfield  {journal} {\bibinfo
  {journal} {Phys.~Rev.~Lett.}\ }\textbf {\bibinfo {volume} {126}},\ \bibinfo
  {pages} {074501} (\bibinfo {year} {2021}{\natexlab{a}})}\BibitemShut
  {NoStop}%
\bibitem [{\citenamefont {Buaria}\ \emph
  {et~al.}(2021{\natexlab{b}})\citenamefont {Buaria}, \citenamefont {Clay},
  \citenamefont {Sreenivasan},\ and\ \citenamefont {Yeung}}]{BCSY21b}%
  \BibitemOpen
  \bibfield  {author} {\bibinfo {author} {\bibfnamefont {D.}~\bibnamefont
  {Buaria}}, \bibinfo {author} {\bibfnamefont {M.~P.}\ \bibnamefont {Clay}},
  \bibinfo {author} {\bibfnamefont {K.~R.}\ \bibnamefont {Sreenivasan}}, \ and\
  \bibinfo {author} {\bibfnamefont {P.~K.}\ \bibnamefont {Yeung}},\ }\bibfield
  {title} {\enquote {\bibinfo {title} {Small-scale isotropy and ramp-cliff
  structures in scalar turbulence},}\ }\href@noop {} {\bibfield  {journal}
  {\bibinfo  {journal} {Phys.~Rev.~Lett.}\ }\textbf {\bibinfo {volume} {126}},\
  \bibinfo {pages} {034504} (\bibinfo {year} {2021}{\natexlab{b}})}\BibitemShut
  {NoStop}%
\bibitem [{\citenamefont {Siggia}(1981)}]{Siggia:1981a}%
  \BibitemOpen
  \bibfield  {author} {\bibinfo {author} {\bibfnamefont {E.~D.}\ \bibnamefont
  {Siggia}},\ }\bibfield  {title} {\enquote {\bibinfo {title} {Numerical study
  of small-scale intermittency in three-dimensional turbulence},}\ }\href@noop
  {} {\bibfield  {journal} {\bibinfo  {journal} {J. Fluid Mech.}\ }\textbf
  {\bibinfo {volume} {107}},\ \bibinfo {pages} {375--406} (\bibinfo {year}
  {1981})}\BibitemShut {NoStop}%
\bibitem [{\citenamefont {Nelkin}(1997)}]{nelkin1999}%
  \BibitemOpen
  \bibfield  {author} {\bibinfo {author} {\bibfnamefont {M}~\bibnamefont
  {Nelkin}},\ }\bibfield  {title} {\enquote {\bibinfo {title} {Enstrophy and
  dissipation must have the same exponents in the high {Reynolds} number limit
  of fluid turbulence},}\ }\href@noop {} {\bibfield  {journal} {\bibinfo
  {journal} {Phys. Fluids}\ }\textbf {\bibinfo {volume} {11}},\ \bibinfo
  {pages} {2202--2204} (\bibinfo {year} {1997})}\BibitemShut {NoStop}%
\bibitem [{\citenamefont {Zeff}\ \emph {et~al.}(2003)\citenamefont {Zeff},
  \citenamefont {Lanterman}, \citenamefont {McAllister}, \citenamefont {Roy},
  \citenamefont {Kostelich},\ and\ \citenamefont {Lathrop}}]{zeff:2003}%
  \BibitemOpen
  \bibfield  {author} {\bibinfo {author} {\bibfnamefont {B.~W.}\ \bibnamefont
  {Zeff}}, \bibinfo {author} {\bibfnamefont {D.~D.}\ \bibnamefont {Lanterman}},
  \bibinfo {author} {\bibfnamefont {R.}~\bibnamefont {McAllister}}, \bibinfo
  {author} {\bibfnamefont {R.}~\bibnamefont {Roy}}, \bibinfo {author}
  {\bibfnamefont {E.~H.}\ \bibnamefont {Kostelich}}, \ and\ \bibinfo {author}
  {\bibfnamefont {D.~P.}\ \bibnamefont {Lathrop}},\ }\bibfield  {title}
  {\enquote {\bibinfo {title} {Measuring intense rotation and dissipation in
  turbulent flows},}\ }\href@noop {} {\bibfield  {journal} {\bibinfo  {journal}
  {Nature}\ }\textbf {\bibinfo {volume} {421}},\ \bibinfo {pages} {146--149}
  (\bibinfo {year} {2003})}\BibitemShut {NoStop}%
\bibitem [{\citenamefont {Buaria}\ \emph {et~al.}(2019)\citenamefont {Buaria},
  \citenamefont {Pumir}, \citenamefont {Bodenschatz},\ and\ \citenamefont
  {Yeung}}]{BPBY2019}%
  \BibitemOpen
  \bibfield  {author} {\bibinfo {author} {\bibfnamefont {D.}~\bibnamefont
  {Buaria}}, \bibinfo {author} {\bibfnamefont {A.}~\bibnamefont {Pumir}},
  \bibinfo {author} {\bibfnamefont {E.}~\bibnamefont {Bodenschatz}}, \ and\
  \bibinfo {author} {\bibfnamefont {P.~K.}\ \bibnamefont {Yeung}},\ }\bibfield
  {title} {\enquote {\bibinfo {title} {Extreme velocity gradients in turbulent
  flows},}\ }\href@noop {} {\bibfield  {journal} {\bibinfo  {journal} {New
  J.~Phys.}\ }\textbf {\bibinfo {volume} {21}},\ \bibinfo {pages} {043004}
  (\bibinfo {year} {2019})}\BibitemShut {NoStop}%
\bibitem [{\citenamefont {Buaria}\ and\ \citenamefont {Pumir}(2022)}]{BP2022}%
  \BibitemOpen
  \bibfield  {author} {\bibinfo {author} {\bibfnamefont {D.}~\bibnamefont
  {Buaria}}\ and\ \bibinfo {author} {\bibfnamefont {A.}~\bibnamefont {Pumir}},\
  }\bibfield  {title} {\enquote {\bibinfo {title} {Vorticity-strain rate
  dynamics and the smallest scales of turbulence},}\ }\href@noop {} {\bibfield
  {journal} {\bibinfo  {journal} {Phys.~Rev.~Lett.}\ }\textbf {\bibinfo
  {volume} {128}},\ \bibinfo {pages} {094501} (\bibinfo {year}
  {2022})}\BibitemShut {NoStop}%
\bibitem [{\citenamefont {Sreenivasan}\ and\ \citenamefont
  {Kailasnath}(1993)}]{sreeni93}%
  \BibitemOpen
  \bibfield  {author} {\bibinfo {author} {\bibfnamefont {K.~R.}\ \bibnamefont
  {Sreenivasan}}\ and\ \bibinfo {author} {\bibfnamefont {P.}~\bibnamefont
  {Kailasnath}},\ }\bibfield  {title} {\enquote {\bibinfo {title} {An update on
  the intermittency exponent in turbulence},}\ }\href@noop {} {\bibfield
  {journal} {\bibinfo  {journal} {Phys. Fluids A: Fluid Dynamics}\ }\textbf
  {\bibinfo {volume} {5}},\ \bibinfo {pages} {512--514} (\bibinfo {year}
  {1993})}\BibitemShut {NoStop}%
\bibitem [{\citenamefont {Prasad}\ \emph {et~al.}(1988)\citenamefont {Prasad},
  \citenamefont {Meneveau},\ and\ \citenamefont {Sreenivasan}}]{prasad88}%
  \BibitemOpen
  \bibfield  {author} {\bibinfo {author} {\bibfnamefont {R.~R.}\ \bibnamefont
  {Prasad}}, \bibinfo {author} {\bibfnamefont {C.}~\bibnamefont {Meneveau}}, \
  and\ \bibinfo {author} {\bibfnamefont {K.~R.}\ \bibnamefont {Sreenivasan}},\
  }\bibfield  {title} {\enquote {\bibinfo {title} {Multifractal nature of the
  dissipation field of passive scalars in fully turbulent flows},}\ }\href@noop
  {} {\bibfield  {journal} {\bibinfo  {journal} {Phys.~Rev.~Lett.}\ }\textbf
  {\bibinfo {volume} {61}},\ \bibinfo {pages} {74} (\bibinfo {year}
  {1988})}\BibitemShut {NoStop}%
\bibitem [{\citenamefont {Ishihara}\ \emph {et~al.}(2009)\citenamefont
  {Ishihara}, \citenamefont {Gotoh},\ and\ \citenamefont
  {Kaneda}}]{Ishihara09}%
  \BibitemOpen
  \bibfield  {author} {\bibinfo {author} {\bibfnamefont {T.}~\bibnamefont
  {Ishihara}}, \bibinfo {author} {\bibfnamefont {T.}~\bibnamefont {Gotoh}}, \
  and\ \bibinfo {author} {\bibfnamefont {Y.}~\bibnamefont {Kaneda}},\
  }\bibfield  {title} {\enquote {\bibinfo {title} {Study of high-{Reynolds}
  number isotropic turbulence by direct numerical simulations},}\ }\href@noop
  {} {\bibfield  {journal} {\bibinfo  {journal} {Ann.~Rev.~Fluid~Mech.}\
  }\textbf {\bibinfo {volume} {41}},\ \bibinfo {pages} {165--80} (\bibinfo
  {year} {2009})}\BibitemShut {NoStop}%
\bibitem [{\citenamefont {Rogallo}(1981)}]{Rogallo}%
  \BibitemOpen
  \bibfield  {author} {\bibinfo {author} {\bibfnamefont {R.~S.}\ \bibnamefont
  {Rogallo}},\ }\bibfield  {title} {\enquote {\bibinfo {title} {Numerical
  experiments in homogeneous turbulence},}\ }\href@noop {} {\bibfield
  {journal} {\bibinfo  {journal} {NASA Technical Memo}\ }\textbf {\bibinfo
  {volume} {81315}} (\bibinfo {year} {1981})}\BibitemShut {NoStop}%
\bibitem [{\citenamefont {Buaria}\ and\ \citenamefont
  {Sreenivasan}(2020)}]{BS2020}%
  \BibitemOpen
  \bibfield  {author} {\bibinfo {author} {\bibfnamefont {D.}~\bibnamefont
  {Buaria}}\ and\ \bibinfo {author} {\bibfnamefont {K.~R.}\ \bibnamefont
  {Sreenivasan}},\ }\bibfield  {title} {\enquote {\bibinfo {title} {Dissipation
  range of the energy spectrum in high {Reynolds} number turbulence},}\
  }\href@noop {} {\bibfield  {journal} {\bibinfo  {journal}
  {Phys.~Rev.~Fluids}\ }\textbf {\bibinfo {volume} {5}},\ \bibinfo {pages}
  {092601(R)} (\bibinfo {year} {2020})}\BibitemShut {NoStop}%
\bibitem [{\citenamefont {Buaria}\ \emph
  {et~al.}(2020{\natexlab{a}})\citenamefont {Buaria}, \citenamefont
  {Bodenschatz},\ and\ \citenamefont {Pumir}}]{BBP2020}%
  \BibitemOpen
  \bibfield  {author} {\bibinfo {author} {\bibfnamefont {D.}~\bibnamefont
  {Buaria}}, \bibinfo {author} {\bibfnamefont {E.}~\bibnamefont {Bodenschatz}},
  \ and\ \bibinfo {author} {\bibfnamefont {A.}~\bibnamefont {Pumir}},\
  }\bibfield  {title} {\enquote {\bibinfo {title} {Vortex stretching and
  enstrophy production in high {Reynolds} number turbulence},}\ }\href@noop {}
  {\bibfield  {journal} {\bibinfo  {journal} {Phys.~Rev.~Fluids}\ }\textbf
  {\bibinfo {volume} {5}},\ \bibinfo {pages} {104602} (\bibinfo {year}
  {2020}{\natexlab{a}})}\BibitemShut {NoStop}%
\bibitem [{\citenamefont {Buaria}\ \emph {et~al.}(2022)\citenamefont {Buaria},
  \citenamefont {Pumir},\ and\ \citenamefont {Bodenschatz}}]{BPB2022}%
  \BibitemOpen
  \bibfield  {author} {\bibinfo {author} {\bibfnamefont {D.}~\bibnamefont
  {Buaria}}, \bibinfo {author} {\bibfnamefont {A.}~\bibnamefont {Pumir}}, \
  and\ \bibinfo {author} {\bibfnamefont {E.}~\bibnamefont {Bodenschatz}},\
  }\bibfield  {title} {\enquote {\bibinfo {title} {Generation of intense
  dissipation in high {Reynolds} number turbulence},}\ }\href@noop {}
  {\bibfield  {journal} {\bibinfo  {journal} {Philos. Trans. R. Soc. A}\
  }\textbf {\bibinfo {volume} {380}},\ \bibinfo {pages} {20210088} (\bibinfo
  {year} {2022})}\BibitemShut {NoStop}%
\bibitem [{\citenamefont {Buaria}\ and\ \citenamefont
  {Sreenivasan}(2022)}]{BS2022}%
  \BibitemOpen
  \bibfield  {author} {\bibinfo {author} {\bibfnamefont {D.}~\bibnamefont
  {Buaria}}\ and\ \bibinfo {author} {\bibfnamefont {K.~R.}\ \bibnamefont
  {Sreenivasan}},\ }\bibfield  {title} {\enquote {\bibinfo {title} {Scaling of
  acceleration statistics in high reynolds number turbulence},}\ }\href@noop {}
  {\bibfield  {journal} {\bibinfo  {journal} {arXiv:2202.09682}\ } (\bibinfo
  {year} {2022})}\BibitemShut {NoStop}%
\bibitem [{\citenamefont {Overholt}\ and\ \citenamefont
  {Pope}(1996)}]{overholt96}%
  \BibitemOpen
  \bibfield  {author} {\bibinfo {author} {\bibfnamefont {M.~R.}\ \bibnamefont
  {Overholt}}\ and\ \bibinfo {author} {\bibfnamefont {S.~B.}\ \bibnamefont
  {Pope}},\ }\bibfield  {title} {\enquote {\bibinfo {title} {Direct numerical
  simulation of a passive scalar with imposed mean gradient in isotropic
  turbulence},}\ }\href@noop {} {\bibfield  {journal} {\bibinfo  {journal}
  {Phys. Fluids}\ }\textbf {\bibinfo {volume} {8}},\ \bibinfo {pages} {3128}
  (\bibinfo {year} {1996})}\BibitemShut {NoStop}%
\bibitem [{\citenamefont {Clay}\ \emph
  {et~al.}(2017{\natexlab{a}})\citenamefont {Clay}, \citenamefont {Buaria},
  \citenamefont {Gotoh},\ and\ \citenamefont {Yeung}}]{clay.cpc1}%
  \BibitemOpen
  \bibfield  {author} {\bibinfo {author} {\bibfnamefont {M.~P.}\ \bibnamefont
  {Clay}}, \bibinfo {author} {\bibfnamefont {D.}~\bibnamefont {Buaria}},
  \bibinfo {author} {\bibfnamefont {T.}~\bibnamefont {Gotoh}}, \ and\ \bibinfo
  {author} {\bibfnamefont {P.~K.}\ \bibnamefont {Yeung}},\ }\bibfield  {title}
  {\enquote {\bibinfo {title} {A dual communicator and dual grid-resolution
  algorithm for petascale simulations of turbulent mixing at high {Schmidt}
  number},}\ }\href@noop {} {\bibfield  {journal} {\bibinfo  {journal} {Comput.
  Phys. Commun.}\ }\textbf {\bibinfo {volume} {219}},\ \bibinfo {pages}
  {313--328} (\bibinfo {year} {2017}{\natexlab{a}})}\BibitemShut {NoStop}%
\bibitem [{\citenamefont {Clay}\ \emph {et~al.}(2018)\citenamefont {Clay},
  \citenamefont {Buaria}, \citenamefont {Yeung},\ and\ \citenamefont
  {Gotoh}}]{clay.cpc2}%
  \BibitemOpen
  \bibfield  {author} {\bibinfo {author} {\bibfnamefont {M.~P.}\ \bibnamefont
  {Clay}}, \bibinfo {author} {\bibfnamefont {D.}~\bibnamefont {Buaria}},
  \bibinfo {author} {\bibfnamefont {P.~K.}\ \bibnamefont {Yeung}}, \ and\
  \bibinfo {author} {\bibfnamefont {T.}~\bibnamefont {Gotoh}},\ }\bibfield
  {title} {\enquote {\bibinfo {title} {{GPU} acceleration of a petascale
  application for turbulent mixing at high {Schmidt} number using {OpenMP}
  4.5},}\ }\href@noop {} {\bibfield  {journal} {\bibinfo  {journal} {Comput.
  Phys. Commun.}\ }\textbf {\bibinfo {volume} {228}},\ \bibinfo {pages}
  {100--114} (\bibinfo {year} {2018})}\BibitemShut {NoStop}%
\bibitem [{\citenamefont {Clay}\ \emph
  {et~al.}(2017{\natexlab{b}})\citenamefont {Clay}, \citenamefont {Buaria},\
  and\ \citenamefont {Yeung}}]{clay.omp}%
  \BibitemOpen
  \bibfield  {author} {\bibinfo {author} {\bibfnamefont {M.~P.}\ \bibnamefont
  {Clay}}, \bibinfo {author} {\bibfnamefont {D.}~\bibnamefont {Buaria}}, \ and\
  \bibinfo {author} {\bibfnamefont {P.~K.}\ \bibnamefont {Yeung}},\ }\bibfield
  {title} {\enquote {\bibinfo {title} {Improving scalability and accelerating
  petascale turbulence simulation using {OpenMP}},}\ }in\ \href@noop {} {\emph
  {\bibinfo {booktitle} {Proceedings of {OpenMP} Conference}}}\ (\bibinfo
  {address} {Stony Brook University, NY},\ \bibinfo {year} {2017})\BibitemShut
  {NoStop}%
\bibitem [{\citenamefont {Batchelor}(1959)}]{batch1959a}%
  \BibitemOpen
  \bibfield  {author} {\bibinfo {author} {\bibfnamefont {G.~K.}\ \bibnamefont
  {Batchelor}},\ }\bibfield  {title} {\enquote {\bibinfo {title} {Small-scale
  variation of convected quantities like temperature in turbulent fluid .1.
  {G}eneral discussion and the case of small conductivity},}\ }\href@noop {}
  {\bibfield  {journal} {\bibinfo  {journal} {J. Fluid Mech.}\ }\textbf
  {\bibinfo {volume} {{5}}},\ \bibinfo {pages} {{113--133}} (\bibinfo {year}
  {1959})}\BibitemShut {NoStop}%
\bibitem [{\citenamefont {Pope}(2000)}]{popebook}%
  \BibitemOpen
  \bibfield  {author} {\bibinfo {author} {\bibfnamefont {S.~B.}\ \bibnamefont
  {Pope}},\ }\href@noop {} {\emph {\bibinfo {title} {Turbulent Flows}}}\
  (\bibinfo  {publisher} {Cambridge University Press},\ \bibinfo {year}
  {2000})\BibitemShut {NoStop}%
\bibitem [{\citenamefont {Buaria}\ \emph
  {et~al.}(2020{\natexlab{b}})\citenamefont {Buaria}, \citenamefont {Pumir},\
  and\ \citenamefont {Bodenschatz}}]{BPB2020}%
  \BibitemOpen
  \bibfield  {author} {\bibinfo {author} {\bibfnamefont {D.}~\bibnamefont
  {Buaria}}, \bibinfo {author} {\bibfnamefont {A.}~\bibnamefont {Pumir}}, \
  and\ \bibinfo {author} {\bibfnamefont {E.}~\bibnamefont {Bodenschatz}},\
  }\bibfield  {title} {\enquote {\bibinfo {title} {Self-attenuation of extreme
  events in {Navier-Stokes} turbulence},}\ }\href@noop {} {\bibfield  {journal}
  {\bibinfo  {journal} {Nat. Commun.}\ }\textbf {\bibinfo {volume} {11}},\
  \bibinfo {pages} {5852} (\bibinfo {year} {2020}{\natexlab{b}})}\BibitemShut
  {NoStop}%
\bibitem [{\citenamefont {Buaria}\ and\ \citenamefont {Pumir}(2021)}]{BP2021}%
  \BibitemOpen
  \bibfield  {author} {\bibinfo {author} {\bibfnamefont {D.}~\bibnamefont
  {Buaria}}\ and\ \bibinfo {author} {\bibfnamefont {A.}~\bibnamefont {Pumir}},\
  }\bibfield  {title} {\enquote {\bibinfo {title} {Nonlocal amplification of
  intense vorticity in turbulent flows},}\ }\href@noop {} {\bibfield  {journal}
  {\bibinfo  {journal} {Phys.~Rev.~Research}\ }\textbf {\bibinfo {volume}
  {3}},\ \bibinfo {pages} {L042020} (\bibinfo {year} {2021})}\BibitemShut
  {NoStop}%
\bibitem [{\citenamefont {Grossmann}\ \emph {et~al.}(1997)\citenamefont
  {Grossmann}, \citenamefont {Lohse},\ and\ \citenamefont
  {Reeh}}]{grossmann1997}%
  \BibitemOpen
  \bibfield  {author} {\bibinfo {author} {\bibfnamefont {S.}~\bibnamefont
  {Grossmann}}, \bibinfo {author} {\bibfnamefont {D.}~\bibnamefont {Lohse}}, \
  and\ \bibinfo {author} {\bibfnamefont {A.}~\bibnamefont {Reeh}},\ }\bibfield
  {title} {\enquote {\bibinfo {title} {Different intermittency for longitudinal
  and transversal turbulent fluctuations},}\ }\href@noop {} {\bibfield
  {journal} {\bibinfo  {journal} {Phys. fluids}\ }\textbf {\bibinfo {volume}
  {9}},\ \bibinfo {pages} {3817} (\bibinfo {year} {1997})}\BibitemShut
  {NoStop}%
\end{thebibliography}

%

\end{document}